\documentclass[11pt]{article}
\usepackage{latexsym}

\def\ket#1{\mid~\!\!\!{#1}~\!\!\rangle}
\def\bra#1{\langle~\!\!{#1}~\!\!\!\mid}
\def\+-{\buildrel + \over -}

\def\QM{quantum mechanics }
\def\qm{quantum mechanics}
\def\Q{quantum }

\def\QMl{quantum-mechanical }

\def\${\enskip$}
\def\WF{wavefunction }
\def\wf{wavefunction}
\def\M{measurement }
\def\m{measurement}

\def\e{ensemble}
\def\IS{individual-system }

\begin{document}

{\large \bf \noindent On Historical Background to the Ontic Breakthrough.I\\ Polemic Defense of Quantum Reality}\\

\begin{quote}
{\noindent\small\rm Fedor Herbut}\\
{\noindent\scriptsize\it Serbian Academy
of Sciences and Arts, Belgrade, Serbia}\\

{\footnotesize \noindent After stating the author's ontic position, a collage of relevant thoughts of some distinguished foundationally-minded physicists are quoted and polemically commented upon. Thus, a kind of historical background of the recent ontic breakthrough is given. The brealthrough itself is presented in Part II.}\\

\end{quote}

 \normalsize \rm
\section{INTRODUCTION}

Since I'll begin the polemic first Part of this two-part article on the question of reality of the \WF by outlining my position, let me stipulate what I believe.

(i) the world that is accessible to our senses and described by classical (meaning all pre-quantum) physics is real, i. e. it exists outside our consciousnesses in some way completely independently of our observation.

(ii) Quantum experiments are mostly performed on \e s. Quantum-mechanical probability predictions based on the preparation are always confirmed in the \m s. Hence the \e wise reality of the \Q world is beyond doubt. There is no need to discuss these two kinds of reality.

This point of view was put very well by Lyre \cite{Lyre}:

"It is more than obvious that disbelief in the reality of our laboratories, measuring devices
and pointer states does not even get the scientific enterprize off the ground."

The question remains to what extent are the {\bf individual} \Q systems like molecules, atoms, electrons, protons, neutrons neutrinos elementary particles etc. well described by \qm , and what about their reality? A separate and more specific question concerns the reality of the basic concept of \qm , the \WF (the term is used slightly incorrectly as a popular synonym for an abstract \Q pure state or state vector).

These two questions on reality allow and require different beliefs, different arguments, and different investigations, In the present article, in sections 2 and 3, it is explained why the author says 'yes' to both questions in the title. Further on, in sections 4-?, a polemic presentation of opinions  and arguments of a number of foundationally concerned physicists is given. Sections 4-? treat the reality problem of the \Q world; and section ? discuss  the reality of the \wf .\\

According to my own point of view, \QM and the wavefunction are 'knowledge' about the \Q world that exist out there independently of us and our observations, i. e., putting it philosophically, they are primarily epistemic entities. But they are also, in some way, real 'out there', or ontic entities. Every prediction that can be drawn from the \WF and that can be tested on ensembles does have also an individual meaning of some kind. I have elaborated this tenet on the important concept of the individual-system meaning of indeterminate \Q properties in terms of 'degrees of presence' \cite{FHStud}. More will be stated in subsection 2.1 in the second Part.\\

 Further details on my views are presented in sections 2 and 3.\\

\section{Ensembles and individual systems}

When pondering  what the \WF can say about an individual \Q system, we run into trouble. Gleason has shown \cite{Gleason} that any general \Q state (density operator), in particular any pure state (\WF ) is equivalent to the entirety of probability distributions on the spectra of all possible \Q observables. This is a firm theoretical foundation of the \e wise reality of the \wf , but it leaves no ground for the individual system.\\

Nevertheless, a relevant point for our search for \IS reality of the \WF is the following {\bf ensemble paradox}. Let a wavefunction \$\Psi\$ predict that if one measures an observable \$O\$ its eigenvalue \$o_k\$ is to be obtained with certainty (probability one). {\it Version one} of the paradox: I carry out the \m , and I obtain some other result. Does this contradict the ensemblewise prediction of \$\Psi\$? It does not, because the empirical basis of statistical prediction are infinite ensembles. In particular, if I perform an unlimited number of \m s, the relative frequency of the result \$o_k\$ will be one.  But nothing is said about the relative frequency
in a finite ensemble. We know from mathematics that an infinite number sequence can tend to one though it has any number of crazy figures at the beginning. It is how it behaves in the infinity that matters.

By now it must be clear to the reader what are {\it versions two} and three of this paradox. I may spend my whole life repeating the \M in question and never obtain the result \$o_k\$ predicted with certainty. Still I am not in contradiction with \$\Psi\$. In {\it version three}, all mankind as long as civilization lasts, may it be millions of years, may repeat the \M and never obtain the predicted result. And still one is not in contradiction with the prediction of \$\Psi\$. Any finite ensemble is, actually, irrelevant. And it is only finite ensembles that are accessible to us.

Nature could play tricks on physicists, but it does not. The finite ensembles that we are dealing with in the laboratory are well behaved. Why so?

I suggest the following answer: It is because \$\Psi\$ does describe also the individual system; in other words, it has individual-system reality. Hence the systems in a finite ensemble cannot behave substantially differently than those  in the infinity; they are all described by the same wavefunction.

In the next section a simple conceptual framework of an approach to description of reality, which I believe to be the basis of all scientific research, is expounded.\\

\section{Improved-approximation approach to description of reality}

Every child knows the circle, and when it is taught to evaluate its circumference from the diameter, it encounters the mystical number \$\pi\$. Many of us have forgotten how baffled we were when we faced an irrational for the first time. It is easy with the rationals. It begins with, e. g., the Indian chieftain saying "I'll give three horses for your daughter", and he shows three fingers etc. Then comes total induction and division and the rationals are operationally covered. Nobody wonders if they exist; of course they do. But do irrationals exist in nature?

They are hard-to-understand  (unattainable) limits of infinite sequences of rationals. The sequences on the real axis may come from below, from above, or from both sides of the irrational. This diversity of 'approaches' to the irrational via rationals increases in the plane, and becomes really plentiful in the complex infinite-dimensional Hilbert space of quantum mechanics.

In first approximation the relation between the circumference and the diameter of the circle is 3.1 . Equivalently, the value of \$\pi\$ lies somewhere in the half-closed interval [3.1,3.2). In the widely known
second approximation \$\pi\$ is  3.14, or equivalently, within [3.14,3.15), etc. In principle, one can improve on this in as many steps as one likes. But one cannot reach the exact value; equivalently, one cannot narrow down the interval to one point of the real axis.

My contention is that much of \Q 'reality' (or \Q ontology) has an analogous conceptual position to that of \$\pi\$ on the real axis, but the 'approaches' to it in physics are far more abundant than the mentioned 'approaches' to \$\pi\$ by rationals. The essence of the mystery of the transcendental (or unattainable) is already contained in \$\pi\$. Concrete examples discussed below may illustrate this.

In the advocated conceptual framework for the relation between epistemology and ontology in \QM I lean rather on mathematics than on philosophy. The \Q theories, the 'bricks' of which are the wavefunctions, constitute a partially ordered set, i. e., a set in which for some pairs of elements one can say which is an improvement on the other. Naturally, transitivity is valid.

I'l give  some examples. In theoretical nuclear physics \cite{Preston} it is known that the nucleon theory is an improvement on the proton-neutron one, because the former {\bf contains} the latter, i. r., it reproduces the proton and the neutron states of the nucleon in terms of  the third component of isotopic spin. Besides, the nucleon theory has tangible {\bf advantages} in describing nuclear structure and nuclear processes.

In elementary-particle physics one has further improvement: the so-called barion octet. (One can look it up in Wikipedia.) It deals with \Q numbers additional to isotopic spin as the barion number, strangeness etc.\\

Another example is the composition of matter and particles: all matter consists of molecules, these are made up of atoms, which contain nuclei and electrons; the nuclei consist of protons and neutrons, these are made up of quarks, and who knows what further steps of 'breaking up particles' the future will bring.

As an example of pairs of theories that cannot be compared for improvement is the single-particle model and the model of nuclear matter in nuclear physics. They apply to different aspects of nuclei, and none of them is an improvement on the other.\\

In a branch of \QMl research, one allows for the possibility that distinct observers describe a given individual \QMl system by distinct \wf s. Here the epistemic side prevails in the sense that each of these \wf s describes correctly some aspects of the given individual system, gives correct probability distributions on some observables, and the selections of these observables differ. The distinct \wf s must be compatible, i. e., they must be able to describe one and the same \Q system. Next, it is desirable to pool the useful information contained in them to obtain an improved \wf . This research, to which I have made a modest contribution \cite{FHpooling} (see the references therein), is a good example of the improved-approximation approach.\\

Let me now turn to opinions of           other physicists.\\

\section{ZEH}

I'll begin with some thoughts of Zeh \cite{Zeh1} (beginning of the Abstract in the arxiv version; omitted in the book). I believe that Zeh's position is close to mine.

\begin{quote}
{\bf Z1:} "Schr¨odinger's wave function shows many aspects of a state of incomplete knowledge or information ("bit"): (1) it is usually defined on a space of classical configurations, (2) its generic entanglement is, therefore, analogous to statistical correlations, and (3) it determines probabilities of measurement outcomes. Nonetheless, quantum superpositions (such as represented by a wave function) define individual physical states ("it"). This
conceptual dilemma may have its origin in the conventional operational foundation
of physical concepts, successful in classical physics, but inappropriate
in quantum theory because of the existence of mutually exclusive operations
(used for the definition of concepts). In contrast, a hypothetical realism,
based on concepts that are justified only by their universal and consistent
applicability, favors the wave function as a description of (thus nonlocal) physical reality."
\end{quote}

The "It or Bit?" question will become
the question of ontic or epistemic nature of the \WF in the second decade of our century. (See Part II.)

One should note the "{\it or}" in the question. As for my part, I am not against the {\it Bit}, let alone the {\it It}; I am against the "or" if it is   understood in the exclusive sense.

If the answer were strictly just "It", then one would suppress the fact that \qm , as every natural science, is our knowledge about the world outside our consciousness. If, on the other hand, "Bit" by itself carried the day in its extreme form, then our contents of consciousness would be all that there is with a false reference to non-existing reality like in our dreams. "Bit" in its non-extreme form might leave us confused not knowing  how much reality is there in our knowledge.

It seems to me that the answer should be: {\bf It and Bit}. There is a real \Q world, independent of our cognition, 'out there', the "It". Its basic brick is the reality of the wave function. On the other hand, here are we, with our consciousnesses,  observing and manipulating the outside reality. This is the "Bit". Philosophers might say: ontology parallelled by epistemology.

The "operational foundation" mentioned by Zeh appears to be analogous to my "rationals" (cf the preceding section) handled by us operationally, while the "irrationals" (\$\pi\$ as an example) may correspond to Zeh's
"concepts that are justified only by their universal and consistent applicability".\\

 Zeh, in a letter \cite{Zeh2} among other things, says:

\begin{quote}
{\bf Z2:} "Mohrhoff insists on an intrinsic "fuzziness" of nature. However, this fuzziness (just as the related concepts of uncertainty or complementarity) is the consequence of insisting on classical kinematical concepts. In my opinion it indicates that these concepts are just wrong (that is, inappropriate for fundamental purposes) - unless you prefer Bohm's theory. The wave function can always and consistently be assumed to be exactly defined - regardless of what we happen to know. It is in this sense that I define it to be "real"."
\end{quote}

I believe that Zeh meant the wave function of a sufficiently encompassing system, perhaps the whole universe \cite{Zeh3}. On account of all-pervading entanglement, more or less all subsystems of this large system are not described by wavefunctions. They are described by (reduced) density operators the physical meaning of which is 'improper mixtures' \cite{D'Esp1}.\\

In slight disagreement with Zeh, I am not sure that all \Q reality is describable by \wf s (and the mentioned reduced density operators that are derivable from them). For example, the center of mass (of a system of particles) requires a point position. It is a delta-function that does not belong to the Hilbert space. But it can be the limes of \wf s (if one of the 5 topologies in the Hilbert space is suitably made use of). I am sure there are also other examples. The point to note is that all description of reality is somehow determined by \wf s (in my opinion).\\

Another point in Zeh's views that I find interesting and important is lack of a need to search for a sub\Q reality (as one gets the impression reading his reaction \cite{Zeh4} to the article on \Q physics without observer by Goldstein et al. \cite{Gold}). In other words, this standpoint seems to imply that the state of an individual \Q system is described by a \WF and nothing else. (In Part II a great deal will be written about a hypothetical sub\Q world.) Zeh stresses in his reaction that the sub\Q entities are fictitious, inaccessible operationally. It seems to me that one cannot know for sure that what is inaccessible today will not become accessible tomorrow. It is a belief.\\

Two more thoughts of Zeh (on the anti-realism of Niels Bohr) are given at the end of section 8.

About the confirmation of Zeh's (and my) belief in the reality of the \WF by the BPR theorem see end of subsection 2.2 in Part II.\\

\section{WIGNER}

Wigner \cite{Wigner} says (on p. 382):

\begin{quote}
{\bf Wi1:} "... the fact,
undeniable in my opinion, that our
impressions form the primitive
reality."
\end{quote}

On the preceding page {\it ibid.}
Wigner says:

\begin{quote}
{\bf Wi2:} "... even though
'reality' in the physical and
particularly microphysical world
may be a questionable concept, one
cannot accept the observations of
another person, and the resulting
content of his consciousness, to
be less real than those of
ourselves. This suggests that one
treat all observations, undertaken
by oneself or another person, on a
more nearly equal basis."
\end{quote}

The present author concurs that
his observations and those of
others are 'real', but I "cannot accept"
(to use Wigner's expression) that {\bf what
one observes} is of "questionable"
reality. I think, one should make
all three of the mentioned steps:
My impressions are real, so are
those of others, and so are the
objects of our impressions, i. e.,
the outside world, the reality of
which rests on the reality of the
wave function.\\

The position of the present author
is that there is a quantum reality 'out
there', which in some way implies classical reality, and that the wavefunction
is the basic brick. The idea leans on
the reality concept championed by
Einstein \cite{Einstein} who said (on p. 60):

\begin{quote}
{\bf E:} "The belief in an
external world independent of the
perceiving object is the basis of
all natural science."\\
\end{quote}

\section{ADLER}

C. G. Adler \cite{Adler} (p. 878, right column) begins his pondering the idea of realism by giving it a definition.

\begin{quote}
{\bf Adl1:} "Realism, all
would seem to agree, involves the
belief that there is an underlying
reality that we are striving
literally to describe."
\end{quote}

I do not agree completely. Namely, this "describing" reality, towards which one is striving according to Adler, can't always be done "literally"
because we cannot compare our
knowledge with the unknown
"underlying reality". "Literally"
should be replaced by
"approximately as well as
possible" because, all that we can possibly do
is to improve on our knowledge, striving to come closer to reality (as I have elaborated my view in section 3).

There may be counterparts of rationals among our scientific ideas, but we must take great pains to prove it. Adler's discussion of the neutrino below teaches us about this convincingly.\\

Adler further discusses the
experimental verification of the
existence of the neutrino, and he
articulates his scepticism
concerning quantum reality
({\it ibid.}, p. 880, left column,
penultimate passage).

\begin{quote}
{\bf Adl2:} "The neutrino is
then a necessary constituent of
the theory and the associated
experiment. It exists as a
building block of physics, but
does it necessarily exist apart
from the physics that defines it -
I think not!"
\end{quote}

This appears to be a perfectly
correct argument, except that the conclusion does not follow. The criticism is carried too far in it.

One may disagree with Adler's answer to his question "does it necessarily exist apart
from the physics that defines it". The 'theory' (physics in this case)
can be regarded as our {\it path
towards reality} like the mentioned approximating rationals are for \$\pi\$. There might exist
other paths, other "physics that
defines it", but the entity 'out
there' towards which they lead,
can be one and the same. At least,
so it is assumed in \Q realism. Adler's example does not contradict this assumption, nor makes it less likely.

Nobody asks if \$\pi\$ exists "apart from" a given convergent series of rationals that may define it because, as it is usual in powerful mathematics, one can prove that all mentioned paths lead to the same goal, which is \$\pi\$. Admittedly, the cognitive
paths leading towards the concept of the neutrino are far more complex, but the basic idea may still be the same as in the case of the irrational. Adler's pessimistic conclusion "I think not" is far from necessary. \\

Adler continues his train of
thoughts ({\it ibid.}, p. 880,
right column, passage -3):

\begin{quote}
{\bf Adl3:} "In a real sense
the experimenter, like an artist,
manufactured the neutrino."\\
\end{quote}

We do "manufacture" our representations, but their objects are 'out there' in full reality. Like we "manufacture" a convergent infinite sequence of rationals that determine \$\pi\$. But this manufacturing would be of little value if \$\pi\$ did not exist (independently of the particular converging sequence).\\

Further, Adler says ({\it ibid.}, p.
881, left column, passage -3):
\begin{quote}
{\bf Adl4:} "... we as
physicists never know if we have
chosen {\it the right}
explanation. In fact, we have no
reason to believe that there is a
right explanation."
\end{quote}

I suggest to understand these
words as saying that there is no
unique way towards reality. One
must agree with him about the
non-uniqueness, but not with his
doubts about reality. At the end
of the passage, Adler says:

\begin{quote}
{\bf Adl5:} "Viewed in this
way, it might be better to say
that physics is {\it arealistic}
(italics by F. H.) rather than
anti-realistic (that is, "without
making a claim of realism" as
opposed to "claiming realism is
wrong")."
\end{quote}

From my point of view, Adler's "arealistic" position is far more acceptable than an "anti-realistic" one. But it still is not quite acceptable. One should uphold Einstein's leading idea that science has the aim to describe, I would say 'as well as it can', the reality 'out there'.\\

Further,  Adler says ({\it ibid.},
p. 881, right column, about the
middle):
\begin{quote}
{\bf Adl6:} "The question of
the neutrino's independent
existence is unimportant. If the
idea works, and we agree that it
works, that's enough. This is the
position I think most of us as
physicists almost unknowingly
take. And by and large we are
comfortable with it."
\end{quote}

These words call to mind the well known infamous witty saying "Shut up and calculate". I agree that this seems to be the (regrettable) position (knowingly or unknowingly) of most of the quantum physicists.

There is a saying that the ostrich sticks his head in the sand in order not to see any danger. (Animal behaviorists might not agree; but this is a good metaphor for illusionary escapism.)

As to the last sentence in the quote, I am comfortable with \$3.14\$ for \$\pi\$ when I do not want to get involved with the marvel of irrationals.\\

\section{Duhem and Harr\'e}

The quoted warning of Adler is supported by
the philosopher Duhem \cite{Duhem}, who (on p. 187) wrote:

\begin{quote}
{\bf Du:} "Physics is not a
machine which lets itself to be
taken apart; we cannot try each
piece in isolation ... (it) is a
system that must be taken as a
whole; it is an organism."
\end{quote}

This is the famous Duhemian holism: the claim that no scientific hypothesis can be tested in isolation by some {\it experimentum crucis}, but only theories as a whole.

In spite of the general validity of Duhemian holism for physics, the wavefunction can be considered to be comparable to a machine (in the Duhemian sense). Namely, the wavefunction not only implies the probabilities of all possible measurement results, but it is also uniquely determined by the totality of these probabilities (\cite{Gleason}). In this sense the wavefunction "lets itself to be taken apart" into probabilities of individual events (or results of measurement).

Hence, if the wavefunction describing an individual \Q system is real, then the probabilities, which are its 'pieces', must be real too. They are 'degrees of presence' of the \Q system in the event at issue (as in delocalization) (cf \cite{FHStud}).\\

Adler, in a desire to
find support for his argument,
quotes another author
Harr\'{e} \cite{Harre}, who says:

\begin{quote}
{\bf H:} "Realism seems
suited to sciences like anatomy,
physiology, and chemistry, while
physics seems to be the archetypal
positivist or phenomenalist
science in its more fundamental
reaches. This is because at the
boundaries of knowledge the
phenomena are the limits
of the conceivable."\\
\end{quote}

To my mind, the "limits of the
conceivable" "at the boundaries of
knowledge" just means that, in
quantum physics, we have to expand
our imagination and improve our intuition to be able to improve on our approximation to reality. More precisely, we have to put in more effort to find the required wave functions, and, perhaps more importantly, we have to stretch our imagination to understand the knowledge on reality that the wave functions contain.\\

Finally, let us go back to Adler and to his remark inspired by Harr\'e.

(\cite{Adler}, p. 882, left column, first passage in his Conclusion):

\begin{quote}
{\bf Adl7:} "The point to be
made is that when physics works
(as Harr\'{e} says) "at the limits
of the conceivable" it blurs the
distinction between discovery
and creation."\\
\end{quote}

I feel that Adler's word
"creation" in this case should be understood
as "expanding our imagination" as I
have put it above.

My insisting on analogy with \$\pi\$ may sound as a vulgarization, but I personally feel to be "at the limits of the conceivable" already when I see a long string of integers after \$3.\$ (an improvement on \$3.14\$), and when I am told that it is too short, and that it is doomed to be always too short however much one works to prolong it. Already this does "blur the distinction between discovery and creation" for me. But it does not shake my belief that \$\pi\$ exists 'out there' and that it is well defined in mathematics. Unfortunately, in physics many concepts are not so well defined.

I have argued as convincingly as I could
that, in spite of these sobering
warnings of Adler, Duhem and Harr\'e, the belief in the
reality of the neutrino. or more generally, in the  \Q objects of
investigation in physics, and of
the wave function of individual \Q systems in particular,
can, and I think should, be firmly held.\\

\section{Vaihinger, Niels Bohr,  D'Espagnat, and Zeh}

Adler's point of view (in the preceding sections) bears some resemblance to that of the neo-Kantian philosopher  Vaihinger \cite{Vaihinger} (p.5, left column, bottom), who had argued that the underlying reality of the world remains \textbf{unknowable}, but we behave \textbf{''as if''} the constructions of physics such as electrons, protons and electro-magnetic waves exist, and to this extent such 'heuristic fictions' constitute our reality.

Apparently, Vaihinger admits the existence of a reality 'out there' (that what is "unknowable"). I would like to understand his claim that it is 'unknowable'  in the sense of denying literal or exact knowledge (in my simile with \$\pi\$, that it is not as a rational, that it is not attainable by finite operations that we are able to perform).

Vaihinger also argued that we act ''as if'' other minds exist as well, and take this to be part of psychic reality. This fact seems to suggest that Vaihinger's scepticism about our ability to recognize what is 'out there' (even outside our personal mind) might even be deeper than the transcendental unattainability of irrationals. The latter can be reproduced with arbitrarily good precision. Vaihinger's ontology may be thought to be separated by a unbridgeable gap from even the limit of our subjective 'heuristic fictions'.

Then we are doomed to Plato's cave, seeing only shadows on the wall of the cave unable to see the beings that make them outside the cave because we cannot leave the cave. This is a pessimism concerning our worldview that I personally do not share.

Unfortunately, in \QM anti-realism goes back to the great founder, Niels Bohr, himself \cite{Bohr}:

\begin{quote}
{\bf Bo:} "There is no quantum
world. There is only an abstract
quantum physical description. It is
wrong to think that the task of physics
is to find out how nature is. Physics
concerns what we can say about nature."
\end{quote}

These words of the great physicist shocked me, and motivated my careful studies of intriguing and sophisticated experiments \cite{FHOuMandel}, \cite{FHScully1}, \cite{FHScully2}. I believe now firmly that Bohr's words were not right if taken literally.
I believe that there is a quantum world. It does require an abstract description. But this is hardly surprising.

I cannot put my view better than Fagundes \cite{Fagundes} did.

\begin{quote}
{\bf F:} "... physics progresses by increasing degrees of abstraction. This is only natural since 'concrete' ideas are just those of our too limited ordinary sense experience."
\end{quote}

The present author believes that \QM does describe physical reality in spite of the fact that it is "abstract".

The great \QMl analyst B. d'Espagnat gave recently \cite{D'Esp2} an in-depth discussion of \Q reality. In the very abstract he says:

\begin{quote}
{\bf D'Esp1:} "Contrary to classical physics, which was strongly objective i.e. could be interpreted
as a description of mind-independent reality, standard quantum mechanics
is only weakly objective, that is to say, its statements, though inter-subjectively valid, still merely refer to operations of the mind.".
\end{quote}

A few lines later he adds:

\begin{quote}
{\bf D'Esp2:} "It is shown however that this does not rule out a broader form of realism, called here 'open realism', restoring the notion of mind-independent reality.
\end{quote}

Let me close my disapproval of the anti-realism of Bohr by quoting two remarks by Zeh \cite{Zeh1} (last lines in second section):

\begin{quote}
{\bf Z3:} "... none of these great physicists was ready to dismiss the condition that reality must be local (that is, defined in space and time). It is this requirement that led Niels Bohr to abandon microscopic reality completely (while he preserved this concept for the classical realm of events)."
\end{quote}

{\it Ibid.}, in a further section, under the witty title "That ITsy BITsy Wave Function" Zeh writes the following:

\begin{quote}
{\bf Z4:} "The investigation of quantum objects thus required various, mutually
incompatible, operational means. This led to incompatible (or "complementary")
concepts, seemingly in conflict with a microscopic reality. Niels
Bohr's ingenuity allowed him to recognize this situation very early. Unfortunately,
his enormous influence (together with the dogma that the concept
of reality must be confined to objects in space and time) seems to have
prevented his contemporaries to explain it in terms of a more general (nonlocal)
concept that is successfully used but not directly accessible by means
of operations: the universal wave function."
\end{quote}

Let us for a moment part with the insight of the discoverer of decoherence \cite{Zeh4}, a major break-through in the foundations of modern \Q physics, and turn to a completely opposite approach.\\

\section{Fuchs, Schack, and Peres}

In a recent article \cite{Fuchs} the authors present an intricate argument of handling \Q coherence and lack of it in the Bayesian way, and they say (in their Abstract):
\begin{quote}
{\bf F+S:} "Bringing the argument to the context of quantum measurement theory, we show that "quantum decoherence" can be understood in
purely personalist terms — quantum decoherence (as supposed in a von Neumann
chain) is not a physical process at all, but an application of the reflection principle.
From this point of view, the decoherence theory of Zeh, Zurek, and others as a
story of quantum measurement has the plot turned exactly backward."
\end{quote}

I cannot even argue against this because I feel that it is in this article where "the plot is turned exactly backward". I do have respect for the Bayesian approach \cite{Bayes}, \cite{RMPreview}, but I can accept it only as a systematic way of  working out the epistemic part of \Q physics, but not playing down and belittling the ontological part, which is, to my mind, the unavoidable object of all epistemic endeavors.\\

In another, well known, article \cite{FP} Fuchs and Peres argue that \QM does not need interpretation. In their view the alleged interpretations only burden the \Q theory with unnecessary philosophical additions. They convey the impression that the \Q formalism is simple, consistent, and in itself easily understood.

What they offer to substantiate these claims
is "a series of radical philosophic statements regarding the meaning of the wave function" - point out Dennis and Norsen in an incisive criticism \cite{DenNors} of the Fuchs-Peres article. Then they go on quoting from the criticized article. The wave function "is only a mathematical expression for evaluating probabilities"; "no wave function exists either before or after we conduct an experiment"; "the wave function is not an objective entity"; "the time dependence of the wave function does not represent an evolution of a physical system" etc.

Fuchs and Peres contend that the \Q probabilities refer only to macroscopic events. Dennis and Norsen level against this idea the following argument. "Where exactly is the cut between micro- and macroscopic, and why should such a cut enter the fundamental laws of physics? If one electron is not objectively real, and two electrons are not objectively real, why should a collection of \$10^{23}\$ electrons be real? Things like temperature and elasticity may be emergent properties, but surely {\it existence as such} is not. Nothing real can emerge from that which doesn't exist."

Fuchs and the late Asher Peres, as eminent physicists, admit that "the possible existence of an objective reality, independent of what observers perceive"
cannot be excluded in principle (the "hedge" in their article) . But even if reality exists, they argue, it is unimportant in practice for "using the theory and understanding its nature".\\

Hardy and Spekkens take a different view \cite{Hardy} in their short article with the title "Why Physics Needs Quantum Foundations":

\begin{quote}
{\bf Ha-Sp:} "...There is no question that quantum theory works well as a tool for predicting what will occur
in experiments. But just as understanding how to drive an automobile is different from understanding how it works or how to fix it should it break down, so too
is there a difference between understanding how to use quantum theory and understanding what it means. The
field of quantum foundations seeks to achieve such an understanding. In particular, it seeks to determine the
correct interpretation of the formalism. It also seeks to determine the principles that underlie quantum theory...."
\end{quote}

It is time to find some support for the idea of \Q realism.\\

\section{MacPhee and Schlegel}

The position MacPhee   and Schlegel  take in their discussion \cite{MacPhee} is close to my views:

\begin{quote}
{\bf M:} "But even if the
really real always escapes him,
progress in physics means pinning
it down more and more closely."
\end{quote}

Immediately after this, Schlegel
comments \cite{Schlegel}:

\begin{quote}
{\bf S:} I believe, in
suggesting that physics moves
toward depictions of reality that
contain more truth, as each law or
theory is replaced by one of
greater breadth and adequacy. The
development from the crude atomic
theory of Democritus  (which in a
rough way is still true) to the
present level of sophisticated
information about atomic structure
is an obvious case in point."\\
\end{quote}

After this explicit support, I'll turn to a firmly opposed opinion.\\

\section{Hartle}

Hartle \cite{Hartle} says:

\begin{quote}
{\bf Ha1:} "...the state
will be called an objective
property if an assertion of what
the state is can be verified by \m
s on the individual system without
knowledge of the systems previous
history."
\end{quote}

In the second passage after this,
he claims:
\begin{quote}
{\bf Ha2:} "In \QM there are
no \m s on an individual system
which can determine its state."
\end{quote}

At the end of the passage he draws
from this the conclusion:

\begin{quote}
{\bf Ha3:} "A \QMl state is,
therefore, {\bf not} an objective
property of the individual
system."
\end{quote}

More than 40 years have passed
since Hartle had these critical
thoughts. In the meantime
brilliant Israeli physicists
invented theoretically so-called
protective \m (see the review \cite{protect.meas} and the references therein), which might
seem to do precisely what Hartle
was after as a criterion for
'objective property' of the wave
function.

Even if we disregard protective \m , which may not be quite what Hartle had in mind, his criterion for objectivity seems to be {\it inappropriately  restrictive}. Though we may think of the wave function as describing the reality of an individual system in the experiment, the experimental results are always statistical.

If we give up disallowing "previous history" (which may be preparation) required by Hartle, and we assume that we know \$\Psi\$ due to knowledge of the preparation, which is the maximally favorable case, and we perform an individual measurement of the observable \$\ket{\Psi}\bra{\Psi}\$ and obtain the result \$1\$, we cannot say that we have verified that the individual system is in the state \$\Psi\$. This is so because we might obtain the same result for any other wave function \$\Phi\$ that is not orthogonal to \$\Psi\$.

Quantum experimental verification rests on the fact that \$\Psi\$ describes both the individual system and the ensemble of equally prepared individuals. Hence,
measurement can be performed in a \textbf{subensemble}. Then the {\it complementary subensemble} is "without disturbance", and the wave function that the ensemble represents empirically can thus be established.

In the spirit of the EPR \cite{EPR} criterion of reality, it can be said
that what can be observed without disturbance must already have been present or actual. Since, as mentioned, the {\it complementary subensemble} is "without disturbance", this should be the proper verification, in principle, in \qm .

The recommended idea of confirmation may appear to be far from Hartle's notions. But it is in accord with the fact that, unlike in classical physics, in \Q physics one cannot, in general, measure without disturbing the system. Quantum observation is not like careful bird watching.

At first sight, one might object to the proposed verification criterion (that should take the place of Hartle's too restricted one) that it verifies the objectivity of the ensemble, not of the individual system. But ensembles consist of individual systems. The former cannot be real without reality of the latter. (This would be a similar fallacy as if one contended that society exists, but individual people do not.)\\

Hartle levels another apparently devastating argument against the reality of the wave function claiming that indeterminate (probability between zero and one) predictions of a wavefunction cannot be reality, only incomplete knowledge about reality. The argument is dealt with in \cite{FHStud} arguing that the mentioned author is guided by classical intuition, which is in this case not suitable for quantum mechanics. The alternative of realistic degrees of presence (in quantum entities) is suggested to be the proper view.\\

\section{More controversy}

Costa de Beauregard, in his article \cite{Costa} (p.1 second passage) gives a short account of Popper's position with the following words "Realism, in his view, does not, and must not yield any
ground because of intelligibility problems raised jointly by theoretical and experimental physics ...". Costa de Beauregard, unfortunately, does not agree with Popper; I do.\\

In this article I have used the very simplifying attempt to understand the reality 'out there' in terms of rationals and irrationals (like \$\pi\$). This may appear confusing since the wave function, containing maximal possible \Q knowledge, should (in some sense) literally describe the relevant quantum properties. Actually, it does so, but often it is a model system of which the wavefunction gives a full description, and this system only approximates the true one. It is just like when a rational stands for an irrational, but, this time, the model system plays the role of the approximate rational. (Cf the discussion of pooling of different compatible \wf s at the end of section 3.)

Most \Q anti-realists seem only to follow the lead of the great Niels Bohr (cf the quotation of Bohr in section 6??). When a great physicist takes, unfortunately,  a wrong turn at a crossroad, then many admirers follow suit.\\

It is likely that realism, just like anti-realism,  cannot be
made conclusive and necessary. Logically, there is room for the opposite belief (cf the "hedge" that Fuchs and Peres had to admit (in section 7??). But for me it is hard to find it plausible that all those ingenious in-depth theories and beautiful intricate laboratory experiments that \Q physicists keep creating are, in the long run, only empty impressions,
not really different from my
\textbf{dreamworld}, where similarly vivid and fanciful impressions can appear,  and where there is no
objective reality underlying the
contents of my consciousness.

Very few, perhaps no one of the mentioned authors above, have doubts about the reality of the classical world. The quote from Harr\'{e} (cf section 5??) testifies to this. But, to repeat the above important argument, can one believe in the reality of bodies made of the order of \$10^{23}\$ atoms, and not in the reality of the building atoms themselves? Isn't it like believing that the 50th store of a skyscraper is there, but it does not really have a ground floor. Or, to use another simile, it is like having a tree with a beautiful, large top, but without roots.\\

I ran into a sentence that Sokal \cite{Sokal} wrote. He was waging a war against some misuses of natural sciences in cultural sciences, and he put his dissatisfaction with anti-realism in a rather drastic manner.
\begin{quote}
{\bf S:} "Anyone who believes that the laws of physics are mere social conventions is invited to try transgressing those conventions from the windows of my apartment. (I live on the twenty-first floor.)"
\end{quote}

I must agree with the despair that Dennis and Norsen \cite{DenNors} utter at the end of their critical article (cf section 7??).
\begin{quote}
{\bf D+N:} "It is frightening that the same anti-realism Sokal ridiculed there can be put forward as a supposedly natural uncontroversial interpretation of \Q theory. It is even more frightening that this anti-realism is glibly passed off as non-philosophical, for this suggests not only that many physicists have accepted a fundamentally anti-scientific set of philosophical ideas but that, in addition, they have done so unwittingly.

Does the community of {\it physicists} - those individuals whose lives are dedicated  to observing, understanding, and learning to control the physical world - really accept as solid, hard-nosed science the idea (hedged or not) that there is no physical world - that, instead, it is all in our minds?"\\
\end{quote}

To the mind of the present author, lack of belief in the reality of the \Q world degrades \Q research to
psychotherapy to cure our inborn
curiosity when, actually, according
to this belief, there is nothing to be curious about.\\

\section{Concluding remarks}

I like to end the polemic part of this essay quoting the words of, as I understand,
a firm believer in reality of \qm . Jaynes \cite{Jaynes} writes:

\begin{quote}
{\bf Ja:} "... it is
pretty clear why present quantum theory not
only does not use - it does not even dare to
mention - the notion of a "real physical
situation". Defenders of the theory say that
this notion is philosophically naive, a
throwback to outmoded ways of thinking, and
that recognition of this constitutes deep new
wisdom about the nature of human knowledge. I
say that it constitutes a violent
irrationality, that somewhere in this theory
the distinction between reality and our
knowledge of reality has become lost, and the
result has more the character of medieval
necromancy than of science. It has been my
hope that quantum optics, with its vast new
technological capability, might be able to
provide the experimental clue that will show
how to resolve these contradictions."
\end{quote}

Isn't it possible that Jaynes' hope is
beginning to come true, at least to some extent, on account of recent research results like in the work of Scully {\it et al.} \cite{Scully1} and Kim {\it et al.} \cite{Scully2}?  Perhaps \QMl insights based on the \wf -reality belief along the lines of the mentioned studies
\cite{FHOuMandel}, \cite{FHScully1} and \cite{FHScully2} of the present author, may help.\\

To use a borrowed metaphor, an article as the present one 'skates rather lightly over the surface of a very deep pool': it does not do justice to the depth of the problems nor to that of the thoughts of the quoted scientists. But breadth and depth are complementary; you can't do both, not in the same article. My choice has been in favor of some kind of breadth.

\end{document}